\documentclass[graybox]{svmult}

\usepackage{mathptmx}       
\usepackage{helvet}         
\usepackage{courier}        
\usepackage{type1cm}        
\usepackage{makeidx}         
\usepackage{graphicx}        
\usepackage{multicol}        
\usepackage[bottom]{footmisc}

\usepackage{latexsym}
\usepackage{amsmath}
\usepackage{amsfonts}
\usepackage{amssymb}
\usepackage{cite}
\usepackage{amsmath,amsfonts,amsbsy}

\makeindex             

\newcommand{\cL}{{\cal L}}
\newcommand{\cP}{{\cal P}}

\begin{document}

\title*{Chern-Simons-like Gravity Theories}
\titlerunning{Chern-Simons-like ...}
\author{Eric Bergshoeff, Olaf Hohm, Wout Merbis, Alasdair J. Routh and Paul K. Townsend}
\authorrunning{Eric Bergshoeff et al.}
\institute{Eric Bergshoeff and Wout Merbis \at Centre for Theoretical Physics, University of Groningen, Nijenborgh 4, 9747 AG Groningen, The Netherlands, \email{e.a.bergshoeff@rug.nl, w.merbis@rug.nl}
\and Olaf Hohm \at Center for Theoretical Physics, Massachusetts Institute for Technology, Cambridge, MA 02139, USA, \email{ohohm@mit.edu}
\and Alasdair J. Routh and Paul K. Townsend \at Department of Applied and Mathematical Physics, Centre for Mathematical Sciences, University of Cambridge, Wilberforce Road, Cambridge, CB3 0WA, U.K.
 \email{A.J.Routh@damtp.cam.ac.uk, P.K.Townsend@damtp.cam.ac.uk}}
%
%
\maketitle

\abstract{A wide class of three-dimensional gravity models can be put into ``Chern-Simons-like'' form. We perform a Hamiltonian analysis of the general model and then specialise to Einstein-Cartan Gravity, General Massive Gravity, the recently proposed Zwei-Dreibein Gravity and a further parity-violating generalisation combining the latter two.
}

\section{Introduction: CS-like gravity theories}
\label{sec:1}

In three space-time dimensions (3D), General Relativity (GR) can be interpreted as a Chern-Simons (CS) gauge theory of the 3D Poincar\'e,  de Sitter (dS) or anti-de Sitter (AdS) group, depending on the value of the cosmological constant \cite{AT,Witten:1988hc}. The action is the integral of a Lagrangian three-form $L$ constructed from the wedge products of Lorentz-vector valued one-form fields:  the dreibein $e^a$ and the dualised spin-connection $\omega^a$. Using a notation in which the wedge product is implicit, and a ``mostly plus'' metric signature convention, we have 
\begin{equation}\label{bergtown:CSGR}
L = -e_a  R^a + \frac{\Lambda}{6} \varepsilon^{abc} e_a e_b e_c \, ,
\end{equation}
where $R^a$ is the dualised Riemann 2-form:
\begin{equation}
R^a= d\omega^a + \frac{1}{2} \varepsilon^{abc} \omega_b \omega_c\, . 
\end{equation}
This action is manifestly local Lorentz invariant,  in addition to its manifest  invariance under diffeomorphisms, which are on-shell equivalent to local translations. The field equations
are zero field strength conditions for the Poincar\'e or (A)dS group.

Strictly speaking, the CS gauge theory is equivalent to 3D GR only if one assumes  invertibility of the dreibein; this is what allows the Einstein field equations to be written as zero field-strength conditions, and it  is one way to see that 3D GR has no local degrees of freedom, and hence no gravitons. However, there are variants of 3D GR that do propagate gravitons.  The simplest of these are 3D ``massive gravity'' theories found by including certain higher-derivative terms in the action\footnote{It is possible, at least in some cases, to take a massless limit but since ``spin'' is not defined for massless 3D particles, one cannot get a theory of  ``massless gravitons''  this way, if by ``graviton'' we mean a particle of spin-2.}.
The best known example  is Topologically Massive Gravity (TMG),  which includes the  parity-violating Lorentz-Chern-Simons term and leads to third-order field equations that propagate a single spin-2 mode \cite{Deser:1981wh}. A more recent example is New Massive Gravity (NMG) which includes certain curvature-squared terms; this leads to parity-preserving fourth-order equations that propagate a parity-doublet of massive spin-2 modes; combining TMG and NMG we get a parity-violating fourth-order  General Massive Gravity (GMG) theory that propagates two massive gravitons, but with different masses \cite{Bergshoeff:2009hq}.

Although GMG is fourth order in derivatives, it is possible to introduce auxiliary tensor fields to get a set of equivalent first-order equations \cite{Hohm:2012vh}; in  this formulation the fields
can all be taken to be Lorentz vector-valued 1-forms, and the action takes a form that is ``CS-like''  in the sense that it  is the integral of a Lagrangian 3-form defined
without an explicit space-time metric (which appears only on the assumption of an invertible dreibein). The general model of this type can be constructed as follows \cite{Hohm:2012vh}. We start from a
collection of $N$ Lorentz-vector valued 1-forms $a^{r \, a} = a_{\mu}^{r \, a} dx^{\mu}$, where $r=1,\dots,N$ is a ``flavour'' index; the generic Lagrangian 3-form constructible from these
1-form fields is
\begin{equation}\label{bergtown:Lgeneral}
L = \frac12 g_{rs}  a^r \cdot da^s + \frac16 f_{rst} a^r \cdot (a^s \times a^t)\,,
\end{equation}
where $g_{rs}$ is a symmetric constant metric on the flavour space which we will require to be invertible, so it can be used  to raise and lower flavour indices, and the coupling constants $f_{rst}$ define a  totally symmetric ``flavour tensor''.  We now use  a 3D-vector algebra notation for Lorentz vectors in which contractions with $\eta_{ab}$ and $\epsilon_{abc}$ are represented by  dots and crosses respectively. The  3-form  (\ref{bergtown:Lgeneral}) is a CS 3-form when the constants
\begin{equation}
f^{ar} {}_{bs \ ct} \equiv \epsilon^a{}_{bc}f^r{}_{st} \quad \& \quad  g_{ar\ bs} \equiv \eta_{ab} g_{rs}
\end{equation}
are, respectively,  the structure constants of a Lie algebra, and a group invariant symmetric tensor on this Lie algebra\footnote{There are CS gauge theories for which the Lagrangian 3-form is {\it not} of the form  (\ref{bergtown:Lgeneral}) because not all of the generators of the Lie algebra of the gauge group are Lorentz vectors. If  we wish the class of  CS gravity theories to be a subclass of the class of   CS-like gravity theories, we should define the latter by a larger  class of 3-form Lagrangians, as in \cite{Hohm:2012vh}, but  (\ref{bergtown:Lgeneral})  will be sufficient for our purposes.}.  For example, with $N=2$ we may choose $a_1^a =e^a$ and $a_2^a =\omega^a$, and then a choice of the flavour metric and coupling
constants that ensures local Lorentz invariance will yield a CS 3-form equivalent, up to field redefinitions, to  (\ref{bergtown:CSGR}).
For $N>2$, we will continue to suppose that $a_1^a=e^a$ and $a_2^a=\omega^a$, and that the flavour metric and coupling constants are such that the action is local Lorentz invariant, but even with this restriction the generic $N>2$ model will be only CS-like. In particular TMG has a CS-like formulation with $N=3$ and both NMG and GMG have CS-like formulations with $N=4$. Since these models have local degrees of freedom they are strictly CS-like, and not CS models.

The generic $N=4$ CS-like gravity model also includes the recently analysed Zwei-Dreibein Gravity (ZDG) \cite{Bergshoeff:2013xma}. This is a parity preserving massive gravity model with the same local degrees of freedom as NMG (two propagating spin-2 modes of equal mass in a maximally-symmetric vacuum background) but has advantages in the context of
the AdS/CFT correspondence since, in contrast to NMG, it leads to a positive central charge for a possible dual CFT at the AdS boundary.  We shall show here that there is a parity-violating extension of ZDG, which we call ``General Zwei-Dreibein Gravity'' (GZDG).

We focus here on the Hamiltonian formulation of CS-like gravity models for a number of reasons. One is  that the CS-like formulation allows us to discuss various 3D massive gravity models as special cases of a generic model, and this formulation is well-adapted to a Hamiltonian analysis. Another is that there are some unusual features of the Hamiltonian formulation of massive gravity models that are clarified by the CS-like formalism.  One great advantage of the Hamiltonian approach is that it allows a determination of the number of local degrees of freedom independently of a linearised approximation (which can give misleading results). In particular,  massive gravity models typically have an additional local degree of freedom, the Boulware-Deser ghost \cite{Boulware:1973my}.  It is known that GMG has no Boulware-Deser ghost, and this is confirmed by its Hamiltonian analysis,  but ZDG does have a Boulware-Deser ghost for generic parameters \cite{Banados:2013fda}, even though it is ghost-free in a linearised approximation. Fortunately, this problem can be avoided by assuming invertibility of a linear combination of the two dreibeine of ZDG. A special case of this assumption imposes a  restriction of the parameters; this point was made in an erratum to \cite{Bergshoeff:2013xma} and here we present a detailed substantiation of it.  We also present a parity-violating CS-like extension of ZDG, and we show that it has the same number of local degrees of freedom as  ZDG. 

\section{Hamiltonian Analysis}

It is straightforward to put the CS-like model defined by (\ref{bergtown:Lgeneral}) into Hamiltonian form. We perform the space-time split
\begin{equation}\label{bergtown:spacetimesplit}
a^{r\,a} = a_0^{r\,a} dt  +  a_i^{r\,a} dx^i \,,
\end{equation}
which leads to the Lagrangian density
\begin{equation}\label{bergtown:gentimedecomp}
\cL = - \frac12  \varepsilon^{ij} g_{rs} a_{i}^r \cdot \dot{a}_{j}^s + a_{0}^r  \cdot \phi_r\, , 
\end{equation}
where $\varepsilon^{ij} = \varepsilon^{0ij}$. The time components of the fields, $a_0^{r\,a}$, become Lagrange multipliers for the primary constraints $\phi_r^a$:
\begin{equation}\label{bergtown:constraints}
\phi_r^a = \varepsilon^{ij} \left(g_{rs} \partial_i a_j^{s\, a} + \frac12 f_{rst} \left( a_{i}{}^s \times a_{j}{}^t \right)^a \right)\,.
\end{equation}
The Hamiltonian density is just the sum of the primary constraints, each with a Lagrange multiplier $a_0^{r\,a}$,
\begin{equation}\label{bergtown:canH}
\mathcal{H} = - \frac12\varepsilon^{ij} g_{rs} a_i^r \cdot \partial_0 a_j^s - \cL = - a_0{}^r \cdot \phi_r\,.
\end{equation}

We must now work out the Poisson brackets of the primary constraints. Then, following Dirac's procedure \cite{Dirac}, we must consider any secondary constraints. 
We consider these two steps in turn.

\subsection{Poisson brackets and the primary constraints}

The Lagrangian is first order in time derivatives, so the Poisson brackets of the canonical variables can be determined by inverting the first term of \eqref{bergtown:gentimedecomp}; this gives
\begin{equation}\label{bergtown:poissonbr}
\left\{ a_{i\, a}^r (x) , a_{j\, b}^s (y) \right\}_{\rm P.B.} = \varepsilon_{ij}g^{rs} \eta_{ab} \delta^{(2)} (x-y)\,.
\end{equation}
Using this result we may calculate the Poisson brackets of the primary constraint functions. It will be convenient to first define the ``smeared'' functions $\phi[\xi]$ associated to the constraints \eqref{bergtown:constraints} by integrating them against a test function $\xi_a^r(x)$ as follows
\begin{equation}\label{bergtown:phi}
\phi[\xi] = \int_{\Sigma} d^2 x \; \xi_a^r(x) \phi_r^a (x) \,,
\end{equation}
where $\Sigma$ is space-like hypersurface.
In general, the functionals $\phi[\xi]$ will not be differentiable, but we can make them so by adding boundary terms. Varying \eqref{bergtown:phi} with respect to the fields $a_{i}{}^s$ gives
\begin{equation}\label{bergtown:gen_varphi}
\delta \phi [\xi] = \int_{\Sigma} d^2 x \; \xi^{r}_{a} \frac{\delta \phi_r^a}{\delta a_{i}^{s\, b}} \delta a_{i}{}^{s\,b} + \int_{\partial \Sigma} dx \; B[\xi, a,\delta a]\,.
\end{equation}
A non-zero $B[\xi,a,\delta a]$ could lead to delta-function singularities in the brackets of the constraint functions. To remove these, we can choose boundary conditions which make $B$ a total variation
\begin{equation}
\int_{\partial \Sigma} dx \; B[\xi,a, \delta a] = - \delta Q[\xi, a]\,.
\end{equation}
We then work with the quantities
\begin{equation}
\varphi[\xi] = \phi[\xi] + Q[\xi, a]\,,
\end{equation}
which have well-defined variations, with no boundary terms. In our case, after  varying $\phi[\xi]$ with respect to the fields $a_{i}{}^{s}$, we find
\begin{equation}\label{bergtown:gen_varbc}
B[\xi, a, \delta a] = \int_{\partial \Sigma} dx^i \xi_a^r g_{rs} \delta a_{i}{}^{s\,a}\,.
\end{equation}

The Poisson brackets of the constraint functions can now be computed by using equation \eqref{bergtown:poissonbr}. They are given by
\begin{align} \label{bergtown:gen_poissonbr}
\left\{ \varphi(\xi) , \varphi(\eta)  \right\}_{\rm P.B.} = & \; \varphi([\xi, \eta]) + \int_{\Sigma} d^2x \; \xi^r_a \eta^s_b \, \cP_{rs}^{ab}
\nonumber \\
& - \int_{\partial \Sigma} dx^i \; \xi^r \cdot \left[g_{rs}  \partial_i \eta^s + f_{rst} (a_{i}{}^s \times \eta^t)   \right]\, ,
\end{align}
where
\begin{equation}
[\xi ,\eta]^t_c  = f_{rs}{}^{t} \epsilon^{ab}{}_{c} \xi^r_a \eta^s_b\, ,
\end{equation}
and
\begin{align}
\cP_{rs}^{ab} & = f^t{}_{q[r} f_{s] pt} \eta^{ab} \Delta^{pq}  +  2f^t{}_{r[s} f_{q]pt} (V^{ab})^{pq}\,, \label{bergtown:Pmat_def} \\[.2truecm]
V_{ab}^{pq} & =  \varepsilon^{ij} a^p_{i\, a} a^q_{j\, b}\,, \hskip 1truecm \Delta^{pq} = \varepsilon^{ij} a_i^p \cdot a_j^q\,. \hskip 1truecm
\end{align}
In general, adopting non-trivial boundary conditions may lead to a (centrally extended) asymptotic symmetry algebra spanned by the first-class constraint functions if the corresponding test functions $\xi_a^r(x)$ are the gauge parameters of boundary condition preserving gauge transformations. Here we will focus on the bulk theory and assume that the test functions $\xi_a^r(x)$ do not give rise to boundary terms in \eqref{bergtown:gen_varphi} and \eqref{bergtown:gen_poissonbr}. 

The consistency conditions guaranteeing time-independence of the primary constraints are
\begin{equation}
\frac{d}{dt} \phi^b_s = \{\mathcal{H}, \phi^b_s \}_{\rm P.B.} \approx - a_{0\,a}^r \cP_{rs}^{ab} \approx 0\,.
\end{equation}
This expression is equivalent to a set of ``integrability conditions'' which can be derived from the equations of motion.
The field equations of \eqref{bergtown:Lgeneral} are
\begin{equation}\label{bergtown:covEOM}
g_{rs}d a^{s\,a} + \frac12 f_{rst} (a^s \times a^t)^a = 0\,.
\end{equation}
Taking the exterior derivative of this equation and using $d^2 = 0$, we get the conditions
\begin{equation}\label{bergtown:Intcon}
f^t{}_{q[r}f_{s]pt}a^{r\,a} a^p \cdot a^q = 0\,.
\end{equation}
Using the space-time decomposition \eqref{bergtown:spacetimesplit} we have
\begin{equation}\label{bergtown:Intconspacetime}
0= f^t{}_{q[r}f_{s]pt}a^{r\,b} a^p \cdot a^q = a_0^{r}{}_a\cP_{rs}^{ab}\,,
\end{equation}
the right hand side being exactly what is required to vanish for time-independence of the primary constraints. These conditions are 3-form equations in which each 3-form necessarily contains a Lagrange multiplier one-form factor, so they could imply that some linear combinations of the Lagrange multipliers is zero.

If the matrix $\cP_{rs}^{ab}$ vanishes identically then  all primary constraints are first-class and there is no constraint  on any Lagrange multiplier.  In this case the model is actually a Chern-Simons theory, that of the Lie algebra with structure constants $\epsilon^a{}_{bc}f^{r}{}_{st}$.  In general, however, $\cP_{rs}^{ab}$ will not vanish and  $\rm{rank}(\cP)$ will be non-zero. We can pick a basis of constraint functions such that  $3N - \rm{rank}(\cP)$ have zero Poisson bracket with all constraints, while the remaining $\rm{rank}(\cP)$ constraint functions have non-zero Poisson brackets amongst themselves.  At this point, it might appear that  the Lagrange multipliers for the latter set of constraints will be set to zero by the conditions
(\ref{bergtown:Intconspacetime}). However, when one of the fields is a dreibein, this may involve setting $e_0{}^a = 0$. This is not acceptable for a theory of gravity, as the dreibein must be invertible! When specifying a model, we must therefore be clear whether we are assuming invertibility of any fields as it affects the Hamiltonian analysis. In general, if we require invertibility of any one-form field  then we may need to impose further, secondary,  constraints.

In other words, the consistency of the primary constraints is equivalent to satisfying the integrability conditions (\ref{bergtown:Intconspacetime}). If some one-form is invertible, then some integrability condition may reduce to a two-form constraint on the canonical variables, which we must add as a secondary constraint in our theory. We now turn to an investigation of these secondary constraints.

\subsection{Secondary constraints}

To be precise, consider for fixed $s$ the expression $f^{t}{}_{q[r}f_{s]pt}a^{r\,a}$. If the sum over $r$ is non-zero for only one value of $r$, say $a^{r a} = f^a$, and $f^a$ is invertible, then the integrability conditions \eqref{bergtown:Intcon} imply that
\begin{equation}
f^{t}{}_{q[r}f_{s]pt} a^p \cdot a^q = 0\,.
\end{equation}
In particular, taking the space-space part of this two-form, we find
\begin{equation}
\varepsilon^{ij} f^{t}{}_{q[r}f_{s]pt} a_i^p \cdot a_j^q = 0\,,
\end{equation}
which depends only on the canonical variables and is therefore a new, secondary, constraint. One invertible field may lead to several constraints if the above equation  holds for multiple values of $s$. The secondary constraints arising in this way\footnote{Here we should issue a warning: a linear combination of invertible one-forms is not in general invertible, so if $f^{t}{}_{q[r}f_{s]pt}a^{r\,a}$ sums over multiple values of $r$ with each corresponding one-form invertible, this does not in general imply a new constraint. } are therefore the inequivalent components of the field space vector $\psi_{s} = f^t{}_{q[r}f_{s]pt} \Delta^{pq}$. Let $M$ be the number of these secondary constraints, and let us write them as
\begin{equation}\label{bergtown:seccon}
\psi_I = f_{I,pq}\Delta^{pq}\, , \quad I=1,\dots,M\, .
\end{equation}
We now have a total of $3N+M$ constraints.

According to Dirac, after finding the secondary constraints we should add them to the Hamiltonian with new Lagrange multipliers \cite{Dirac}. However, in general this can change the field
equations. To see why let us suppose that we have a phase-space action $I[z]$ for some phase space coordinates $z$, and that  the equations of motion imply the constraint
$\phi(z)=0$. If we add this constraint to the action with a Lagrange multiplier $\lambda$ then we get a new action for which the equations of motion are
\begin{equation}
\frac{\delta I}{\delta z} = \lambda\frac{\partial \phi}{\partial z}\, , \qquad \phi(z)=0\, .
\end{equation}
Any solution of the original equations of motions, together with $\lambda=0$, solves these equations, but there may be more solutions for which $\lambda\ne0$. This is precisely what happens for NMG and GMG (although not for TMG) \cite{Hohm:2012vh}; the field equations of these models lead to a (field-dependent) cubic equation for one of the secondary constraint Lagrange multipliers, leading to two possible non-zero solutions for this Lagrange multiplier\footnote{This problem appears to be distinct from the problem of whether the ``Dirac conjecture'' is satisfied, since that concerns the values of Lagrange multipliers of first-class constraints. It may be related to the recently discussed ``sectors'' issue \cite{Dominici:2013lba}.}. In this case, Dirac's procedure  would appear to lead  us to a Hamiltonian formulation of  a theory that is more general than the one we started with (in that its solution space is larger). Perhaps more seriously, adding the secondary constraints to the Hamiltonian will generally lead to a violation of symmetries of the original model.

Because of this problem, we will omit the secondary constraints from the total Hamiltonian. This omission could lead to difficulties. The  first-class constraints are found by consideration of  the matrix of Poisson brackets of {\it all} constraints, so it could happen that some are linear combinations of primary with secondary constraints. We would then have a situation in which not all  first-class constraints are imposed by Lagrange multipliers in the (now restricted) total Hamiltonian, and this would appear to lead to inconsistencies.  Fortunately, this problem does not actually arise for any of the CS-like gravity models that we shall consider, as they satisfy conditions that we now spell out.

The  Poisson brackets of the primary with the secondary constraint functions are
\begin{align}
\label{bergtown:PBsecondary}
\left\{ \phi[\xi], \psi_I \right\}_{\rm P.B.} =
&\, \varepsilon^{ij} \left[  f_{I,rp} \partial_i (\xi^r) \cdot a^p_{j} +  f_{rs}{}^t f_{I, pt} \xi^r \cdot \left(a_{i}^{s} \times a_{j}^{p} \right) \right]\,,
\end{align}
and the Poisson brackets of the secondary constraint functions amongst themselves are
\begin{equation}
\{\psi_I, \psi_J\}_{\rm P.B.} = 4 f_{I,pq}f_{J,rs} \Delta^{pr} g^{qs}\,.
\end{equation}
We now make the following two assumptions, {\it which hold for all our examples}:

\begin{itemize}

\item We assume that  all  Poisson brackets  of secondary constraints with other secondary constraints
vanish on the full constraint surface. It then follows that the total matrix of Poisson brackets of all $3N+M$ constraint functions takes the form
\begin{equation}\label{bergtown:totalP}
\mathbb{P} = \left( \begin{array}{cc} \cP' & -\left\{ \phi, \psi \right\}^T   \\ \left\{ \phi, \psi \right\} & 0 \end{array}\right) \,,
\end{equation}
where $\cP^\prime$ is  the  matrix of Poisson brackets of the $3N$ primary constraints evaluated on the new constraint surface defined by all $3N+M$ constraints.

\item We assume that inclusion of the secondary constraints in the set of all constraints does not  lead to new  first-class constraints. This means that the secondary constraints must all be second-class, and any linear combination of secondary
constraints and the $\rm{rank}(\cP')$ primary constraints with non-vanishing Poisson brackets on the full constraint surface must be second-class.

\end{itemize}

The rank of $\mathbb{P}$, as given in (\ref{bergtown:totalP}),  is the number of its linearly independent columns. By the second assumption, this  is $M$ plus the number of linearly independent columns of
\begin{equation}
\left( \begin{array}{c} \cP'  \\ \left\{ \phi, \psi \right\} \end{array}\right) \,.
\end{equation}
The number of linearly independent columns of this matrix, as for any other matrix,  is the same as the number of linearly independent rows, which by the second assumption is $\rm{rank}(\cP') + M$. The rank of $\mathbb{P}$, and therefore the number of second-class constraints, is then $\rm{rank}(\cP') + 2M$.

In principle one should now check for tertiary constraints. However, in this procedure the invertibility of certain fields will be guaranteed by the secondary constraints. The consistency of the primary constraints under time evolution can be guaranteed by fixing rank($\mathcal{P}'$) of the Lagrange multipliers. The consistency of the secondary constraints under time evolution, $a^r_{0a}\{\phi^a_r,\psi_I\} \approx 0$ can be guaranteed, under the second assumption, by fixing a further $M$ of the Lagrange multipliers. The fact that these $M$ multipliers are distinct from the rank($\mathcal{P}'$) multiplier fixed before follows from the second assumption. The remaining consistency condition, $\{\psi,\psi\} \approx 0$, is guaranteed by the first assumption.

We therefore have $3N$ - rank($\mathcal{P}'$) - $M$ undetermined Lagrange multipliers, corresponding to the $3N$ - rank($\mathcal{P}'$) - $M$ first-class constraints. The remaining rank($\mathcal{P}'$) + 2$M$ constraints are second-class. The dimension
of the physical phase space per space point is the number of
canonical variables $a_i^{ra}$, minus twice the number of
first-class constraints, minus the number of second-class
constraints, or
\begin{equation}
\label{bergtown:dimcounting}
D = 6N - 2 \times \left( 3N - \rm{rank}(\cP') - M \right) - 1 \times \left( \rm{rank}(\cP') + 2M \right) = \rm{rank}(\cP') \,.
\end{equation}

We will now apply this procedure to determine the number of local degrees of freedom of various 3D gravity models with a CS-like formulation.

\section{Specific Examples}
\label{sec:2}

We will now derive the Hamiltonian form of a number of three-dimensional CS-like gravity models of increasing complexity following the above general procedure.

\subsection{Einstein-Cartan Gravity}

To illustrate our formalism we will start by using it to analyse 3D General Relativity with a cosmological constant $\Lambda$, in its first-order Einstein-Cartan form. 
There are two flavours of one-forms: the dreibein, $a^{e\,a} = e^a$, and the dualised spin-connection $a^{\omega \,a}= \omega^a = \frac12\varepsilon^{abc}\omega_{bc}$. The Lagrangian 3-form is that of  (\ref{bergtown:CSGR}). This takes the general form of \eqref{bergtown:Lgeneral}, with the flavour index $r,s,t, \ldots = \omega, e$. The first step is to read off $g_{rs}$ and $f_{rst}$, and for later convenience we also determine the components of the inverse metric $g^{rs}$ and the structure constants with one index raised, $f^{r}{}_{st}$. The non-zero components of these quantities are:
\begin{align}
g_{\omega e} = -1\,, && f_{eee}=  \Lambda\,, && f_{e\omega \omega} = -1\,, \\ \nonumber
g^{\omega e} = -1\,, && f^{\omega}{}_{ee} = - \Lambda\,, && f^{\omega}{}_{\omega \omega} = 1\,, && f^{e}{}_{e \omega} = 1\,.
\end{align}
These constants define a Chern-Simons 3-form, as mentioned in the introduction; the structure constants are $\varepsilon^a{}_{bc}f^{r}{}_{st}$.  This algebra is spanned by the
Hamiltonian constraint functions, which are all first-class.  In three-dimensions, General Relativity, like any Chern-Simons theory, has no local degrees of freedom.

To see how the details  work in our general formalism, we can work out the matrix \eqref{bergtown:Pmat_def} and find that it vanishes. Then, by equation \eqref{bergtown:dimcounting} the dimension of the physical phase space is
\begin{equation}
D = 12 - 2\times 6 = 0\,,
\end{equation}
as expected. Using \eqref{bergtown:gen_poissonbr} we can also verify that
\begin{equation}\nonumber
\{ \phi^a_\omega, \phi^b_{\omega} \}_{\rm P.B.} = \epsilon^{ab}{}_c \, \phi^c_\omega \,, \quad
\{ \phi^a_e, \phi^b_{\omega} \}_{\rm P.B.} = \epsilon^{ab}{}_c \, \phi^c_e \,, \quad
\{ \phi^a_e, \phi^b_e \}_{\rm P.B.} = - \Lambda \epsilon^{ab}{}_c \, \phi^c_{\omega}\,,
\end{equation}
which is the $SO(2,2)$ algebra for $\Lambda < 0$, $SO(3,1)$ for $\Lambda > 0$ and $ISO(2,1)$ for $\Lambda = 0$, as expected.

\subsection{General Massive Gravity}

General Relativity was a very simple application of our general formalism; as a Chern-Simons theory the Poisson brackets of the constraint functions formed a closed algebra, so it did not require our full analysis. We will now turn to a more complicated example, General Massive Gravity (GMG). This theory does have local degrees of freedom; it propagates two massive spin-2 modes at the linear level. It contains two well known theories of 3D massive gravity as limits: Topologically Massive Gravity (TMG) \cite{Deser:1981wh} and New Massive Gravity (NMG) \cite{Bergshoeff:2009hq}. 

We can write the Lagrangian 3-form of GMG in the general form \eqref{bergtown:Lgeneral}. There are four flavours of one-form, $a^{r\,a} = (\omega^a, h^a, e^a, f^a)$, the dualised spin-connection and dreibein and two extra fields $h^a$ and $f^a$, and the Lagrangian 3-form is
\begin{equation}
\begin{split} \label{bergtown:LGMG}
L_{\rm GMG} = & - \sigma e_a R^a + \frac{\Lambda_0}{6} \epsilon^{abc} e_a e_b e_c + h_a T^a + \frac{1}{2\mu} \left[ \omega_a d\omega^a + \frac13 \epsilon^{abc} \omega_a \omega_b \omega_c \right]  \\
& - \frac{1}{m^2} \left[ f_a R^a + \frac12 \epsilon^{abc} e_a f_b f_c  \right]\,,
\end{split}
\end{equation}
where we recall that $R^a$ is the dualised Riemann 2-form. 
The flavour-space metric $g_{rs}$ and the structure constants $f^{r}{}_{st}$ can again be read off:
\begin{align}
g_{\omega e} = -\sigma\,, && g_{eh} = 1\,, && g_{f\omega} = - \frac{1}{m^2}\,, && g_{\omega\omega} = \frac{1}{\mu}\,, \nonumber \\
f_{e\omega \omega} = - \sigma\,, && f_{eh\omega} = 1\,, && f_{eff} = - \frac{1}{m^2}\,, && f_{\omega\omega\omega} = \frac{1}{\mu}\,, \\ \nonumber
f_{eee}= \Lambda_0\,, &&  && f_{\omega\omega f} = - \frac{1}{m^2}\,.
\end{align}
The next step is to work out the integrability conditions \eqref{bergtown:Intcon}. We find three inequivalent 3-form relations,
\begin{equation}
 e^a e \cdot f = 0\,, \quad
 f^a \left(\frac{1}{\mu } e \cdot f + h \cdot e \right)-h^a e \cdot f = 0\,, \quad
  e^a \left( \frac{1}{\mu } e \cdot f + h \cdot e \right) = 0\,. \\
\end{equation}
We will demand that the dreibein, $e^a$, is invertible. Following our general analysis, we find the two secondary constraints
\begin{equation}\label{bergtown:secconGMG}
\psi_1 = \Delta^{eh} = 0\,, \qquad \psi_2 = \Delta^{ef} = 0\,.
\end{equation}
Next, we compute the matrix $\cP^{ab}_{rs}$ as defined in \eqref{bergtown:Pmat_def}. All the $\Delta^{pq}$ terms drop out because of the secondary constraints, and in the basis $(\omega, h, e, f)$ we get
\begin{equation}\label{bergtown:PmatGMG}
(\cP_{ab})_{rs} = \left(
\begin{array}{cccc}
 0 & 0 & 0 & 0 \\
 0 & 0 & V_{ab}^{ef} & -V_{ab}^{ee} \\
 0 & V_{ab}^{fe} & -2 V_{[ab]}^{hf}  +\frac{1}{\mu }V_{ab}^{ff} & V_{ab}^{he}-\frac{1}{\mu }V_{ab}^{fe} \\
 0 & -V_{ab}^{ee} & V_{ab}^{eh}-\frac{1}{\mu }V_{ab}^{ef} & \frac{1}{\mu }V_{ab}^{ee} \\
\end{array} 
\right) \,.
\end{equation}
We must now determine the rank of this matrix at an arbitrary point in space-time. A Mathematica calculation shows that the rank of $\cP$ is 4. To complete the analysis we need the Poisson brackets of the secondary 
constraint functions $\psi_I$ ($I=1,2$) with themselves and with the 
primary constraint
functions. The Poisson bracket $\{\psi_1,\psi_2\}$ is zero on the 
constraint surface, and the Poisson brackets of $\psi_I$ with the 
primary constraint functions are
\begin{equation} \label{bergtown:GMGpsi1}
\begin{split}
\{ \phi[\xi],\psi_1 \}_{\rm P.B.}& = \epsilon^{ij} \bigg[ \partial_i \xi^h \cdot e_j - \xi^h \cdot (\omega_i \times e_j) -\partial_i \xi^e \cdot h_j + \xi^e \cdot (\omega_i \times h_j)  \\
& + \left(\sigma \xi^e + \frac{1}{m^2} \xi^f \right) \cdot (e_i \times f_j) + \left(\sigma \xi^f + \Lambda_0 \xi^e\right) (e_i \times e_j)
\bigg]\,,
\end{split}
\end{equation}
\begin{equation}\label{bergtown:GMGpsi2}
\begin{split}
\{ \phi[\xi],\psi_2 \}_{\rm P.B.}& =  \epsilon^{ij} \bigg[ \partial_i \xi^f \cdot e_j - \xi^f \cdot (\omega_i \times e_j) -\partial_i \xi^e \cdot f_j + \xi^e \cdot (\omega_i \times f_j) \\
&  + \left( m^2 \xi^h -\frac{m^2}{\mu} \xi^f \right) (e_i \times e_j)  + m^2 \xi^e \cdot \left(e_i \times \left( h_j - \frac{1}{\mu} f_j \right) \right)
\bigg]\,.
\end{split}
\end{equation}
The full matrix of Poisson brackets is a $14\times 14$ matrix $\mathbb{P}$ given by
\begin{equation}\label{bergtown:Pbrackets}
\mathbb{P} =
\left( \begin{array}{ccc}
\cP & v_1 & v_2 \\
-v_1^T & 0 & 0 \\
-v_2^T & 0 & 0
\end{array} \right)\,,
\end{equation}
where the $v_I$, $(I = 1,2)$, are column vectors with components
\begin{equation} \label{bergtown:column}
v_I =  \left( \begin{array}{c}
\{\phi^a_{\omega} , \psi_I \}_{\rm P.B.}\\
\{\phi^a_{h} , \psi_I \}_{\rm P.B.}\\
\{\phi^a_{e} , \psi_I \}_{\rm P.B.}\\
\{\phi^a_{f} , \psi_I \}_{\rm P.B.}
\end{array} \right)\,.
\end{equation}
These brackets can be read off from equations \eqref{bergtown:GMGpsi1} and \eqref{bergtown:GMGpsi2}. The vectors \eqref{bergtown:column} are linearly independent from each other and from the columns of $\cP$, so this increases the rank of $\mathbb{P}$ by $4$. The full $(14 \times 14)$ matrix therefore has rank 8, so eight constraints are second-class and the remaining six are first-class. By eqn.~\eqref{bergtown:dimcounting}, the dimension of the physical phase space per space point is
\begin{equation}
D = 24 - 8 - 2 \times 6 = 4 \,.
\end{equation}
This means there are two local degrees of freedom, and we conclude that the non-linear theory has the same degrees of freedom as the linearised theory, two massive states of helicity $\pm 2$.

\subsection{Zwei-Dreibein Gravity}

We now turn our attention to another theory of massive 3D gravity, the recently proposed Zwei-Dreibein Gravity (ZDG) \cite{Bergshoeff:2013xma}. This is a theory of two interacting dreibeine, $e_1^a$ and $e_2^a$, each with a corresponding spin-connection, $\omega_1{}^a$ and $\omega_2{}^a$. It also has a Lagrangian 3-form of our general CS-like form  \eqref{bergtown:Lgeneral}. Like NMG, ZDG preserves parity and has two massive spin-2 degrees of freedom when linearised about a maximally-symmetric vacuum background, but this does not exclude the possibility of additional local degrees of freedom appearing in other backgrounds. In fact, it was shown by  \cite{Banados:2013fda} that the generic ZDG model  does have an additional local degree of freedom, the Boulware-Deser ghost. We will see why this is so, and also how it can be removed by assuming invertibility of a special linear combination of the two dreibeine. 

The Lagrangian 3-form is
\begin{equation}\label{bergtown:Lbimetric}
\begin{split}
L_{\rm ZDG} =  - M_P \bigg\{ &  \sigma e_{1\,a} R_1{}^a + e_{2\,a} R_2{}^a  + \frac{m^2}{6} \epsilon^{abc} \left( \alpha_1 e_{1\,a}  e_{1\, b}   e_{1\, c}  +  \alpha_2   e_{2\,a} e_{2\, b} e_{2\, c}\right) \\
&  - \cL_{12}(e_1,e_2) \bigg\} \,,
\end{split}
\end{equation}
where $R_1{}^a$ and $R_2{}^a$ are the dualised Riemann 2-forms constructed from $\omega_1{}^a$ and $\omega_2{}^a$ respectively, and the interaction Lagrangian 3-form $L_{12}$ is given by
\begin{equation}\label{bergtown:Lint}
 L_{12}(e_1,e_2) =   \frac12 m^2 \epsilon^{abc} \left( \beta_1 e_{1\, a}   e_{1\, b}   e_{2\, c} + \beta_2 e_{1\, a}   e_{2\, b}   e_{2\, c} \right) \,.
\end{equation}
Here $\sigma = \pm 1$ is a sign parameter, $\alpha_1$ and $\alpha_2$ are two dimensionless cosmological parameters and $\beta_1$ and $\beta_2$ are two dimensionless coupling constants. The parameter $m^2$ is a redundant, but convenient, dimensionful parameter. For now these parameters are arbitrary, but we will soon see that some restrictions are necessary.

From \eqref{bergtown:Lbimetric} we can read off the components of $g_{rs}$ and $f_{rst}$. We will ignore the overall factor $M_P$ to simplify the analysis; after this step they become
\begin{eqnarray}\nonumber \label{bergtown:ZDGfieldmetric}
g_{e_1\omega_1} = g_{\omega_1 e_1} = - \sigma  \,, & & g_{e_2\omega_2} = g_{\omega_2 e_2} =  - 1 \,, \\
f_{e_1 \omega_1 \omega_1} =  - \sigma  \,, & & f_{e_2 \omega_2 \omega_2} = - 1  \,, \\
f_{e_1e_1e_2} =  m^2 \beta_1 \,, & & f_{e_1e_2e_2} =  m^2 \beta_2 \,,\nonumber
\\
f_{e_1e_1e_1} = - m^2 \alpha_1 \,, & & f_{e_2e_2e_2} = -  m^2  \alpha_2 \,. \nonumber
\end{eqnarray}
We also work out the inverse metric $g^{rs}$ and the structure constants $f^r{}_{st}$,
\begin{eqnarray}\nonumber
g^{e_1\omega_1} = g^{\omega_1 e_1} = - \frac{1}{\sigma } \,, & & g^{e_2\omega_2} = g^{\omega_2 e_2} =  - 1  \,, \\
f^{\omega_1}{}_{\omega_1 \omega_1} = f^{e_1}{}_{\omega_1 e_1} = 1 \,, & & f^{\omega_2}{}_{\omega_2 \omega_2} = f^{e_2}{}_{\omega_2 e_2} = 1\,, \\
f^{\omega_1}{}_{e_1e_2} = f^{\omega_1}{}_{e_2e_1} = -\frac{ m^2 \beta_1 }{\sigma}  \,, & & f^{\omega_1}{}_{e_2e_2} =  - \frac{ m^2 \beta_2}{\sigma} \,,\nonumber
\\
f^{\omega_1}{}_{e_1e_1} = \frac{m^2 }{\sigma }\alpha_1 \,, & & f^{\omega_2}{}_{e_2e_2} = m^2 \alpha_2 \,, \nonumber \\
f^{\omega_2}{}_{e_1e_2} = f^{\omega_2}{}_{e_2e_1} = - m^2 \beta_2 \,, & & f^{\omega_2}{}_{e_1e_1} = - m^2 \beta_1 \,.\nonumber
\end{eqnarray}
Equipped with these expressions, we can evaluate the $12\times 12$ matrix of Poisson brackets \eqref{bergtown:gen_poissonbr}, in the flavour basis $(\omega_1,\omega_2,e_1,e_2)$
\begin{align}\label{bergtown:Pmat}
\nonumber (\cP_{ab})_{rs} = & \;\; m^2 \eta_{ab} \left(
\begin{array}{cccc}
 0 & 0 & - \beta _1 \Delta^{e_1e_2} & - \beta _2 \Delta^{e_1e_2} \\[.1truecm]
 0 & 0 &  \beta _1 \Delta^{e_1e_2} &  \beta _2 \Delta^{e_1e_2} \\[.1truecm]
  \beta _1 \Delta^{e_1e_2} & - \beta _1 \Delta^{e_1e_2} & 0 &  - \beta_1 \Delta^{\omega_-e_1}- \beta _2 \Delta^{\omega_-e_2} \\[.1truecm]
  \beta _2 \Delta^{e_1e_2} & - \beta _2 \Delta^{e_1e_2} &  \beta_1 \Delta^{\omega_-e_1} +\beta _2 \Delta^{\omega_-e_2} & 0
\end{array}
\right) \nonumber \\[.2truecm]
& + m^2 \beta_1
\left(
\begin{array}{cccc}
 0 & 0 &  V_{ab}^{e_1e_2} & - V_{ab}^{e_1e_1} \\[.1truecm]
 0 & 0 & -V_{ab}^{e_1e_2} & V_{ab}^{e_1e_1}  \\[.1truecm]
  V_{ab}^{e_2e_1} & - V_{ab}^{e_2e_1} & - (V_{[ab]}^{\omega_1e_2}-V_{[ab]}^{\omega_2e_2}) &
   (V_{ab}^{\omega_1e_1} - V_{ab}^{\omega_2e_1}) \\[.1truecm]
 - V_{ab}^{e_1e_1} &  V_{ab}^{e_1e_1} & (V_{ab}^{e_1\omega_1}-V_{ab}^{e_1\omega_2}) & 0 
\end{array}
\right) \\[.2truecm]
& + m^2 \beta_2
\left(
\begin{array}{cccc}
 0 & 0 & V_{ab}^{e_2e_2} & - V_{ab}^{e_2e_1} \\[.1truecm]
 0 & 0 & - V_{ab}^{e_2e_2} &  V_{ab}^{e_2e_1} \\[.1truecm]
 V_{ab}^{e_2e_2} & - V_{ab}^{e_2e_2} & 0 &
  - (V_{ab}^{e_2\omega_1}-V_{ab}^{e_2\omega_2}) \\[.1truecm]
- V_{ab}^{e_1e_2} &  V_{ab}^{e_1e_2} & -( V_{ab}^{\omega_1e_2} -  V_{ab}^{\omega_2e_2}) &
 (V_{[ab]}^{\omega_1e_1}-V_{[ab]}^{\omega_2e_1}) 
\end{array}
\right) \nonumber\,.
\end{align}
Where $\omega_- \equiv \omega_1 - \omega_2$. We determine the rank of this matrix as before using Mathematica, and find it to be 6. This means that there are $12 - 6 = 6$ gauge symmetries in the theory.

To find the secondary constraints we must study the integrability conditions \eqref{bergtown:Intcon}. There are three independent equations
\begin{align}
(\beta_1 e_1{}^a + \beta_2 e_2{}^a ) e_1 \cdot e_2 = 0\,, \label{bergtown:e1e2} \\
e_2{}^a \omega_- \cdot (\beta_1 e_1 + \beta_2 e_2) - \beta_1 \omega_-{}^a e_1 \cdot e_2 = 0\,, \label{bergtown:omega1}  \\
e_1{}^a \omega_- \cdot (\beta_1 e_1 + \beta_2 e_2) + \beta_2 \omega_-{}^a e_1 \cdot e_2 = 0 \,. \label{bergtown:omega2}
\end{align}
Assuming invertibility of both dreibeine, $e_1^a$ and $e_2^a$, is not sufficient to generate a secondary constraint; from \eqref{bergtown:e1e2} we need that $(\beta_1 e_1{}^a + \beta_2 e_2{}^a)$ is invertible. This does not follow from the invertibility of the two separate dreibeine. Without any secondary constraints, the dimension of the physical phase space, using eqn.~\eqref{bergtown:dimcounting}, is 6. This corresponds to 3 local degrees of freedom, one massive graviton and the other presumably a scalar ghost.

We are interested in theories of massive gravity without ghosts, so we must restrict our general model to ensure secondary constraints. By analysing \eqref{bergtown:e1e2}-\eqref{bergtown:omega2} we see that to derive two secondary constraints, we should assume the invertibility of the linear combination $\beta_1 e_1{}^a + \beta_2 e_2{}^a$. A special case of this assumption, where the ZDG parameter space is restricted to $\beta_1\beta_2 = 0$, but one of the $\beta_i$ is non-zero and the corresponding dreibein is assumed to be invertible, was considered in an erratum to \cite{Bergshoeff:2013xma}. We will first analyse this special case in more detail and then move to the generic case.

\subsubsection{The case $\beta_1\beta_2 = 0$}

In the case that we set to zero one of the two parameters $\beta_i$ we may choose, without loss of generality, to set
\begin{equation}
\beta_2=0\, .
\end{equation}
In this case the invertibility of $e_1{}^a$ alone implies the two secondary constraints.
\begin{equation}\label{bergtown:ZDGSeccon}
\psi_1 = \Delta^{e_1e_2}  = 0\,, \qquad  \psi_2 = \Delta^{\omega_- e_1} = 0 \,.
\end{equation}
These constraints and parameter choices cause the first and last matrices in eqn.~\eqref{bergtown:Pmat} to vanish. The remaining matrix $\cP_{rs}^{ab}$ has rank 4.

The secondary constraints \eqref{bergtown:ZDGSeccon} are in involution with each other, and their brackets with the primary constraint functions are given by
\begin{equation} \label{bergtown:ZDGpsi1}
\begin{split}
\{ \phi[\xi], \psi_1 \}_{\rm P.B.} = \varepsilon^{ij} \bigg[ & \partial_i \xi^{e_1} \cdot e_{2\,j} - \xi^{e_1} \cdot \omega_{1\,i} \times e_{2\, j} - \partial_i \xi^{e_2} \cdot e_{1\,j} + \xi^{e_2} \cdot \omega_{2\,i} \times e_{1\,j} \\
& - \left(\xi^{\omega_1} - \xi^{\omega_2} \right) \cdot e_{1\,i} \times e_{2\,j}
\bigg]\,,
\end{split}
\end{equation}
and
\begin{align} \label{bergtown:ZDGpsi2}
\{ \phi[\xi], \psi_2 \}_{\rm P.B.} = & \varepsilon^{ij} \bigg[  ( \partial_i \xi^{\omega_1} - \partial_i \xi^{\omega_2}) \cdot e_{1\,j} - (\xi^{\omega_1} - \xi^{\omega_2}) \cdot  (\omega_{2\,i} \times e_{1\, j})  - \partial_i \xi^{e_1} \cdot \omega_{-\,j} \nonumber \\
& + \xi^{e_1} \cdot (\omega_{1\,i} \times \omega_{-\,j}) + m^2 \left( \sigma \beta_1  \xi^{e_1} + \alpha_2 \xi^{e_2}  \right) \cdot (e_{1\,i} \times e_{2\,j})  \\
& - m^2 \left( (\sigma \alpha_1 + \beta_1) \xi^{e_1} - \sigma \beta_1 \xi^{e_2}  \right) \cdot (e_{1\,i} \times e_{1\,j})
\bigg]\,. \nonumber
\end{align}
The full matrix of Poisson brackets is again a $14\times 14$ matrix $\mathbb{P}$ given by \eqref{bergtown:Pbrackets}, where the $v_I$ with $I = 1,2$ are now
\begin{equation} \label{bergtown:columnZDG}
v_I =  \left( \begin{array}{c}
\{\phi^a_{\omega_1} , \psi_I \}_{\rm P.B.}\\
\{\phi^a_{\omega_2} , \psi_I \}_{\rm P.B.}\\
\{\phi^a_{e_1} , \psi_I \}_{\rm P.B.}\\
\{\phi^a_{e_2} , \psi_I \}_{\rm P.B.}
\end{array} \right)\,.
\end{equation}
These brackets can be read off from equations \eqref{bergtown:ZDGpsi1} and \eqref{bergtown:ZDGpsi2}. The vectors \eqref{bergtown:columnZDG} are linearly independent from each other and with the columns of $\mathbb{P}$, so this increases the rank of $\mathbb{P}$ by $4$. The total number of second-class constraints is 8, leaving 6 first-class constraints. Using \eqref{bergtown:dimcounting} we find that for general values of the parameters $\alpha_1$, $\alpha_2$ and $\beta_1$ the dimension of the physical phase space per space point is 4. This corresponds to the 2 local degrees of freedom of a massive graviton.

\subsubsection{The case of invertible $\beta_1 e_1{}^a + \beta_2 e_2{}^a$}

The more general case is to assume invertibility of the linear combination of the two dreibeine, $\beta_1 e_1{}^a + \beta_2 e_2{}^a$. In this case, to keep track of the invertible field, we make a field redefinition in the original Lagrangian \eqref{bergtown:Lbimetric}. We define, for $\beta_1 + \sigma \beta_2 \neq 0$,
\begin{equation}
e^a = \frac{2}{\beta_1 + \sigma \beta_2} \left( \beta_1 e_1{}^a + \beta_2 e_2{}^a \right)\,, \qquad
f^a = \sigma e_1{}^a - e_2{}^a \,.
\end{equation}
For convenience we will work with the sum and difference of the spin connections\footnote{Note that the sum of the two connections also transforms as a connection, while the difference transforms as a tensor under the diagonal gauge symmetries}
\begin{equation}
\omega^a = \frac12 \left(\omega_1{}^a + \omega_2{}^a \right)\,, \qquad
h^a = \frac12 \left(\omega_1{}^a - \omega_2{}^a \right)\,.
\end{equation}
In terms of these new fields, the ZDG Lagrangian 3-form is
\begin{align}
L = &  - M_P \bigg\{  \sigma e_a R^a(\omega) + c f_a R^a(\omega) + f_a \mathcal{D} h^a + \frac12 \epsilon_{abc} (\sigma e^a + c f^a) h^b h^c \, \nonumber \\
& + m^2 \epsilon_{abc} \left( \frac{a_1}{6} e^a e^b e^c - \frac{b_1}{2} e^a e^b f^c - \frac{b_2}{2} e^a f^b f^c \right. \\
& \qquad \qquad \quad \left. + \frac{(c^2 -1)b_1 - 2\, c\, \sigma b_2}{6} f^a f^b f^c \right) \bigg\}\,, \nonumber
\end{align}
where $\mathcal{D}$ is the covariant derivative with respect to $\omega$. The new dimensionless constants $(a_1, b_1, b_2, c)$ are given in terms of the old $(\alpha_I, \beta_I)$ parameters as follows
\begin{align}
a_1 = & \; \frac{1}{8} \left( \alpha_1 - 3 \sigma \beta_1 - 3 \beta_2 + \sigma \alpha_2 \right)\,, & b_1 = & \; \frac{\alpha_2 \beta_1 + \beta_2^2 - \beta_1^2 - \alpha_1 \beta_2}{4(\beta_1 + \sigma \beta_2)}\,, \\
b_2 = & - \frac{\alpha_1\beta_2^2 + \sigma \beta_1\beta_2^2 + \beta_1^2 \beta_2 + \sigma \alpha_2 \beta_1^2}{2(\beta_1 + \sigma \beta_2)^2}\,, & c = & \;  \frac{\sigma \beta_2 - \beta_1}{\sigma \beta_2 + \beta_1}\,. \nonumber
\end{align}
By construction, this theory has two secondary constraints for invertible $e^a$. Indeed, when we calculate the integrability conditions \eqref{bergtown:Intcon} for this theory we find the three equations
\begin{equation}\label{bergtown:IntconexZDG}
\frac12(\beta_1 + \sigma \beta_2) e^a  f \cdot e =  0\,, \qquad   \frac12 (\beta_1 + \sigma \beta_2)e^a h \cdot e = 0\,, 
\end{equation}
and
\begin{equation} 
\frac12(\beta_1 + \sigma \beta_2) \left( h^a f \cdot e + f^a h \cdot e\right) =  0 \,.
\end{equation}
From \eqref{bergtown:IntconexZDG} we can derive two secondary constraints, since we assumed that $e^a$ was invertible and that $\beta_1 + \sigma \beta_2 \neq 0$. The secondary constraints are
\begin{equation}
\psi_1 = \Delta^{fe} = 0 \,, \qquad \psi_2 = \Delta^{he} = 0\,.
\end{equation}
After imposing these constraints, the matrix of Poisson brackets in the basis $(\omega, h, f, e)$ reduces to
\begin{equation}
 (\cP_{ab})_{rs} = \frac12 m^2 (\beta_1 + \sigma \beta_2) \left(
\begin{array}{cc}
0   &   0   \\
0   &   Q
\end{array} \right)\,,
\end{equation}
where
\begin{equation}\label{bergtown:PmatexZDG}
Q =
\left(
\begin{array}{ccc}
 0 &  V_{ab}^{ee} &  - V_{ab}^{ef} \\
 V_{ab}^{ee} & 0  &  - V_{ab}^{eh} \\
 - V_{ab}^{fe} & - V_{ab}^{he} & V_{[ab]}^{hf} \\
\end{array}
\right) \,. 
\end{equation}
Using the same techniques as previously, we find that this matrix has rank 4.

The secondary constraints are again in involution with themselves, and their brackets with the primary constraint functions are given by
\begin{align} \label{bergtown:psi1PB}
\{ \phi[\xi], \psi_1 \}_{\rm P.B.}  = & \; \varepsilon^{ij} \bigg[ \partial_i \xi^{f} \cdot e_{j} - \xi^{f} \cdot \omega_{i} \times e_{j} - \partial_i \xi^{e} \cdot f_{j} + \xi^{e} \cdot \omega_{i} \times f_{j}  \nonumber \\
& - \left(\sigma \xi^{e} + c\, \xi^{f} \right) \cdot e_{i} \times h_{j}   - \left(c\, \xi^{e} + \sigma (c^2 -1) \xi^{f} \right) \cdot f_{i} \times h_{j} \\
&- \xi^h \cdot \left( \sigma e_i \times e_j + 2 c\, e_i \times f_j + \sigma (c^2 -1 ) f_i \times f_j \right)
\bigg]\,, \nonumber
\end{align}
and
\begin{align} \label{bergtown:psi2PB}
\{ \phi[\xi], \psi_2 \}_{\rm P.B.} = \; & \varepsilon^{ij} \bigg[  \partial_i \xi^{h} \cdot e_{j} - \xi^{h} \cdot \omega_{i} \times e_{j}  - \partial_i \xi^{e} \cdot h_{j}  + \xi^{e} \cdot \omega_{i} \times h_{j} \nonumber \\
&  + m^2 \left( (c\, \sigma a_1 +  b_1) \xi^{e} - (c\, \sigma b_1 -  b_2) \xi^{f}  \right) \cdot e_{i} \times e_{j}  \\ \nonumber
&  -  m^2 \left( (c\, \sigma b_1 - b_2)  \xi^{e}  +  ( (c^2-1) b_1 - c\, \sigma b_2) \xi^{f}  \right) \cdot e_{i} \times f_{j} \\
& - (c\, \xi^e + \sigma (c^2 -1) \xi^f) \cdot h_i \times h_j   - \xi^h \cdot \left(c\, e_i \times h_j + \sigma (c^2-1) f_i \times h_j \right)
\bigg]\,. \nonumber
\end{align}
For generic values of the parameters these constraints increase the rank of the total matrix of Poisson brackets, $\mathbb{P}$, by 4, leading to a $14 \times 14$ matrix of rank 8. This implies that there are eight second-class constraints and six first-class constraints, leading to two degrees of freedom, those of two massive spin-2 modes in 3 dimensions.

To summarize, demanding the presence of secondary constraints in ZDG to remove unwanted degrees of freedom forces us to make an additional assumption about the theory. We must assume invertibility of a linear combination of the two dreibeine. With an additional restriction on the parameter space of the theory, the invertibility of one of the original dreibeine is sufficient to remove the Boulware-Deser ghost. Note that only one dreibein (or one combination of the two dreibeine) need be assumed invertible. This suggests that we identify its square as the ``physical'' metric with which distances are measured.  In the case where $\beta_1\beta_2 = 0$, this suggestion is supported by the fact that the second dreibein may be solved for in terms of the invertible dreibein and its derivatives, leading to an equation of motion for a single dreibein containing an infinite sum of higher derivative contributions \cite{Bergshoeff:2014eca}. It would be interesting to investigate whether this is also possible in the generic case.

\subsection{General Zwei-Dreibein Gravity}

It is natural to look for a parity violating generalisation of ZDG, just as GMG is a parity violating version of NMG. To this end we add to the ghost-free, $\beta_2 = 0$ ZDG Lagrangian 3-form \eqref{bergtown:Lbimetric} a Lorentz-Chern-Simons (LCS) term for the spin-connection $\omega_{1}{}^a$.\footnote{It is also possible to include a LCS term for $\omega_2{}^a$, in this case the expressions presented in this subsection are only slightly modified and lead to the same conclusion.}
\begin{equation}
L_{\rm GZDG} = L_{\rm ZDG}(\beta_2 = 0) + \frac{M_P}{2\mu} \omega_{1\,a} \left( d \omega_1{}^a + \frac13 \epsilon^{abc} \omega_{1\,b} \omega_{1\,c} \right) \,.
\end{equation}
The introduction of the LCS term for $\omega_1{}^a$ introduces non-zero torsion for $e_1{}^a$. One might consider adding a torsion constraint for $e_1{}^a$, enforced by a Lagrange multiplier field $h^a$, but this introduces new degrees of freedom \cite{Bergshoeff:2013xma}. In any case, the equations of motion for General ZDG are such that the torsion constraint is not needed in order to solve for the spin-connections, and there exists a scaling limit similar to the NMG-limit presented in \cite{Bergshoeff:2013xma} where the General ZDG Lagrangian reduces to the GMG Lagrangian \eqref{bergtown:LGMG}.

From the point of view of our general formalism, the addition of the LCS term adds the following non-zero components to $g_{rs}$ and $f_{rst}$
\begin{equation}
g_{\omega_1 \omega_1} = \frac{1}{\mu} \,, \qquad f_{\omega_1 \omega_1 \omega_1} = \frac{1}{\mu}\,.
\end{equation}
The integrability conditions now read
\begin{align}
e_1{}^a  e_1 \cdot e_2 = 0\,, \label{bergtown:e1e2GZDG} \\
e_1{}^a \left( \omega_- \cdot e_1 + \frac{\beta_1 m^2}{\mu} e_1 \cdot e_2 \right) = 0\,, \label{bergtown:omega1GZDG} \\
e_2{}^a \omega_- \cdot e_1  + \left( \frac{\beta_1 m^2}{\mu} e_2{}^a  -  \omega_-{}^a\right) e_1 \cdot e_2 = 0\,. \label{bergtown:omega2GZDG}
\end{align}
Invertibility of $e_1{}^a$ implies the same secondary constraints as in eqn.~\eqref{bergtown:ZDGSeccon}, and the counting of degrees of freedom proceeds analogously. After a linear redefinition of the constraints to $\phi_{\omega'} = \phi_{\omega_1} + \phi_{\omega_2}$, the matrix of Poisson brackets reduces to
\begin{equation}
(\cP_{ab})_{rs} =  m^2 \beta_1 \left(
\begin{array}{cc}
0   &   0   \\
0   &   Q
\end{array} \right)\,,
\end{equation}
where
\begin{equation}\label{bergtown:PmatGZDG}
Q =
\left(
\begin{array}{ccc}
 0 & - V_{ab}^{e_1e_2} &  V_{ab}^{e_1e_1} \\
 - V_{ab}^{e_2e_1} &  -(V_{[ab]}^{\omega_1e_2}-V_{[ab]}^{\omega_2e_2}) +  \frac{\beta_1 m^2}{\mu} V_{ab}^{e_2e_2} &
   (V_{ab}^{\omega_1e_1}-V_{ab}^{\omega_2e_1}) -  \frac{\beta_1 m^2}{\mu} V_{ab}^{e_2e_1} \\
  V_{ab}^{e_1e_1} & ( V_{ab}^{e_1\omega_1} -  V_{ab}^{e_1\omega_2}) -  \frac{\beta_1 m^2}{\mu} V_{ab}^{e_1e_2} &
 \frac{\beta_1 m^2}{\mu} V_{ab}^{e_1e_1}
 \\
\end{array}
\right). 
\end{equation}
We find that this matrix has rank 4. The Poisson brackets of the secondary constraints with the primary ones are now:
\begin{align} \label{bergtown:GZDGpsi1}
\{ \phi[\xi], \psi_1 \}_{\rm P.B.} = \varepsilon^{ij} \bigg[ & \partial_i \xi^{e_1} \cdot e_{2\,j} - \xi^{e_1} \cdot \omega_{1\,i} \times e_{2\, j} - \partial_i \xi^{e_2} \cdot e_{1\,j} + \xi^{e_2} \cdot \omega_{2\,i} \times e_{1\,j}  \nonumber \\
& - \left(\xi^{\omega_1} - \xi^{\omega_2} + \frac{\alpha_1 m^2}{\mu} \xi^{e_1} - \frac{m^2 \beta_1}{\mu} \xi^{e_2} \right) \cdot e_{1\,i} \times e_{2\,j} \\
& \nonumber  + \frac{\beta_1 m^2  }{\mu } \xi^{e_1} \cdot e_{2\,i} \times e_{2\,j}
\bigg]\,,
\end{align}
and
\begin{align} \label{bergtown:GZDGpsi2}
\{ \phi[\xi], \psi_2 \}_{\rm P.B.} = & \varepsilon^{ij} \bigg[  ( \partial_i \xi^{\omega_1} - \partial_i \xi^{\omega_2}) \cdot e_{1\,j} - (\xi^{\omega_1} - \xi^{\omega_2}) \cdot  (\omega_{2\,i} \times e_{1\, j})  - \partial_i \xi^{e_1} \cdot \omega_{-\,j} \nonumber \\
& + \xi^{e_1} \cdot (\omega_{1\,i} \times \omega_{-\,j}) + m^2 \left( \sigma \beta_1  \xi^{e_1} + \alpha_2 \xi^{e_2}  \right) \cdot (e_{1\,i} \times e_{2\,j})  \\
& - m^2 \left( (\sigma \alpha_1 + \beta_1)\xi^{e_1} - \sigma \beta_1 \xi^{e_2}  \right) \cdot (e_{1\,i} \times e_{1\,j}) \nonumber \\
& + m^2 \left( \frac{\alpha_1}{\mu} \xi^{e_1} - \frac{\beta_1}{\mu} \xi^{e_2}\right) \cdot (e_{1\,i} \times \omega_{-\,j}) - m^2 \frac{\beta_1}{\mu} \xi^{e_1} \cdot (e_{2\,i} \times \omega_{-\,j})
\bigg]\,. \nonumber
\end{align}
Again, the secondary constraints are in involution, and the new columns are linearly independent from each other and the original columns. The usual analysis shows that there are 8 second-class constraints and 6 first-class constraints. The total dimension of the physical phase space remains 4, and so the model has the same number of local degrees of freedom as GMG.

\section{Conclusions}

It is a remarkable fact that many of the 3D ``massive gravity''
models that have been found and analysed in recent years have a
CS-like formulation in which the action is an integral over a
Lagrangian 3-form constructed from wedge products of 1-forms  that
include an invertible dreibein. One example not considered here is
Topologically Massive Supergravity \cite{Routh:2013uc}. 

Many of these CS-like models have an alternative formulation as a higher-derivative extension of 3D
General Relativity, and it is certainly not the case that all such higher-derivative extensions can be 
recast as CS-like models. It appears that the unitary (ghost-free) 3D massive models are also special 
in this respect. Whatever the reason may be for this, it is fortunate because the CS-like formalism is well-adapted to a
Hamiltonian analysis, which we have reviewed, and refined, extending the results of \cite{Hohm:2012vh}
for  General Massive Gravity (GMG) to include the recently proposed  Zwei-Dreibein Gravity (ZDG) \cite{Bergshoeff:2013xma}.  

This Hamiltonian analysis leads to a simple determination of the number of local  degrees of freedom, independent of 
any linearisation about a particular background. This allows one to establish that a class of  3D massive gravity models is free of the Boulware-Deser ghost that typically afflicts massive gravity models \cite{Boulware:1973my}. This class includes ZDG, provided a linear combination of the dreibeine is assumed to be invertible.
Conversely, the CS-like formulation of these models can be used as a starting point to find higher-derivative extensions of New Massive Gravity which are
guaranteed to be free of scalar ghosts \cite{Afshar:2014ffa}.

We have also discussed a parity-violating extension of ZDG; it has some similarities to GMG (and has a limit to GMG for a certain 
range of its parameters) so it  could be called ``General Zwei-Dreibein Gravity'' (GZDG). We have shown that it has exactly the same number 
of local degrees of freedom as GMG. We know that ZDG propagates two spin-2 modes of equal mass in a maximally symmetric vacuum,
so it seems that GZDG will propagate two spin-2 modes of different masses, like GMG. It would be interesting to see whether there
is some limit of the parameters of GZDG that sends one mass to infinity keeping the other fixed, because we would then have
a model similar to TMG  but possibly with better behaviour in relation to the AdS/CFT correspondence.

\begin{acknowledgement}
This paper  is based upon lectures given by Eric Bergshoeff and Paul Townsend at the
Seventh Aegean Summer School {\sl Beyond Einstein's Theory of Gravity} in Paros, Greece. E.B., W.M., A.R. and P.T.  thank the organizers of the Paros School for providing an inspiring  environment. We are also grateful to Joaquim Gomis and Marc Henneaux for discussions and correspondence on Hamiltonian methods.
\end{acknowledgement}
%


\providecommand{\href}[2]{#2}\begingroup\raggedright\endgroup


\end{document}